# The Liquid Sheet Shape and Thickness Predictions in Two Impinging Jets Based on Minimum Energy Principle

*A. Kebriaee[1], A. Abdehkakha[2], H. Dolatkhahi[1], S. Kashanj[1]*

[1]*Sharif University of Technology, Tehran, Tehran, Iran, 1458889694*

[2] *University at Buffalo, Buffalo, New York, USA, 14260*

**Abstract**

In an impingement atomizer, a thin sheet of liquid generates by impinging two identical jets. The thickness is one of the most crucial parameters in quantifying the spray characteristics for which several non-unique solutions have been acquired theoretically, so far. Among all, three theoretical solutions presented by Hasson and Peck, Ranz, and Miller are the most reliable expressions in terms of precision. Apparently, all these distinctive solutions cannot be physical simultaneously. However, a theoretical satisfactory explanation for having various non-unique solutions has not been presented. Moreover, a meaningful criterion to recognize the most physical solution is highly desirable. To address this gap, it is hypothesized that an additional constraint should be considered in couple with the continuity and momentum equations to have a unique prediction for the liquid sheet thickness. By employing the minimum energy principle, the minimum stored energy in the liquid sheet is considered the crucial constraint to achieving a unique solution and rejecting non-physical ones. Since the inviscid flow assumption is reasonably valid for the liquid sheet, the minimum energy principle satisfies when the minimum lateral surface predicts. To quantify the lateral surface area the sheet shape is presented in terms of the sheet thickness and numerically calculated by the fourth-order Runge-Kutta method for the three mentioned expressions. Using the liquid shape, the lateral surface areas are calculated and compared in three different impingement angles of 58, 89, and 117 degrees. In addition, the liquid shapes are visually compared with the experimental results obtained by the Shadowgraph technique. In sum, it is proved that Hasson and Peck presented the most physical solution for the liquid sheet thickness.

## Introduction

Impinging jets are widely used in a broad range of applications such as fuel injection, food industry, material and manufacturing, agriculture, etc. [1]. In the liquid propellant engines, the impinging jet atomizer is more desirable compared with the other available types such as airblast [2], gas-liquid [3, 4], and pressure-swirl atomizers [5] because of its simplicity and efficient atomization. As is known, the spray characteristics of an impinging jet are one of the most effective parameters on the performance of a liquid-fueled engine. Therefore, a massive amount of theoretical, experimental, and numerical studies has been performed to identify atomization phases [5-8].

In a typical impinging jet atomizer, the spray formation process could be classified into five main phases; jets impingement, sheet formation, hydrodynamic instability waves propagation, sheet disintegration into ligaments, and finally ligament breakup and atomization. In the first phase, two identical round jets collide at an impingement point. Then, the liquid sheet initiates from the impingement point and stretches in the lateral direction. Thereafter, aerodynamic and hydrodynamic instability waves propagate through the liquid sheet. By growing the instability waves, the sheet disintegrates into ligaments. Finally, the ligaments break into droplets and spray forms [6].

To date, each of these phases has been exclusively characterized. The results showing that the sheet thickness is one of the substantial parameters. In fact, the sheet thickness is directly correlated with the ligament diameter [7] and indirectly with the spray droplets size [8]. That is the reason why most of the experimental [1, 9, 10] and numerical [11, 12] studies are focused on characterizing the liquid sheet thickness and find an expression to accurately predict the liquid sheet thickness. In the sum of all these numerical and experimental studies, the liquid sheet thickness ($h$) mostly depends on the impingement angle of two jets ($2\alpha$), the jet diameter ($d_0$), the velocity of jets ($u_0$), fluid viscosity ($\mu$), and fluid density ($\rho$) [9].

Many theoretical studies have been done so far to put illumination on the physics behind the atomization process. As the main object of all, the required physical assumptions have been discussed to reach an analytical solution for the liquid sheet thickness in a variety of hydrodynamic and geometrical parameters. For the first time, Ranz [13] studied the dynamics of a sheet and reach an analytical solution for the sheet thickness. By following the previous experimental study of Rupe [14], Ranz [13] considered an ideal sheet that is formed by the collision of two identical oblique jets. He assumed the mechanical energy loss is negligible and the kinetic energies of jets would be fully transferred to the surface energy. By applying the mass and momentum equations based on these assumptions, Ranz [13] presented a solution for the film thickness as $h_r = (d_0^2/4r)(1 + 2\cos\alpha\cos\theta)$ where $d_0$ is the jet orifice diameter, $r$ is the radial position from the impingement point, $\alpha$ is the half of impingement angle, and $\theta$ is the angular position. As is stated by Ranz [13], the equation is only valid for the impingement angles higher than $60^o$. Indeed, for the values less than this threshold the Ranz's equation predicted a nonphysical negative value for the sheet thickness.

In the following, Miller [15] used experimental results to prove why Ranz expression is only valid in the impingement angles larger than $60^o$. Miller claimed that Ranz's theory did not capture the backward flow in the sheet. By adding this assumption, he obtained an analytical solution for the sheet thickness as $h_m = d_0^2(\sin\alpha)^2/4r(1 - 2\cos\alpha\cos\theta + \cos^2\alpha)$. He validated this equation with experimental results in all the ranges of impingement angles. later, Hasson and Peck [7] argued the Miller expression because of its low accuracy and presented another theoretical expression as $h_h = d_0^2(\sin\alpha)^3/4r(1 - \cos\alpha\cos\theta)^2$. They assumed the sheet has a constant elliptical shape, the flow is inviscid, and liquid velocity at the sheet remains constant. Their theoretical expression had a good agreement with the Miller and Taylor experimental results [9, 15].

To date, the accuracy and validation of these three expressions have been thoroughly investigated [1, 16-18]. As the outcome of all, it has been proved that Hasson and Peck's expression is the most promising solution that is in close agreement with the experimental results. However, some of Hasson and Peck's theoretical assumptions have been truly argued. Ibrahim and Przekwas [19] asserted that considering an elliptical shape for the sheet is not a valid assumption. Indeed, they claimed the Hasson and peck analysis cannot be used for the sheet shape determination. To make an improvement, they assumed the sheet a free and unbounded shape instead of a constant ellipse but kept the other Hasson and Peck's assumptions such as the constant fluid velocity along the streamline. By assuming the potential flow and neglecting the energy dissipation, they obtained a semi-empirical expression for both the thickness and the shape of the sheet for the low range of Weber numbers.

In the following, Bush and Hasha [17] modified Ibrahim and Przekwas's sheet shape assumption into a thin leaf-shaped sheet that is bounded by a thick stable rim. They showed experimentally and theoretically that describing a sheet in bounded form yields more convincing results. Yang et al. [20] followed the idea presented by Bush and Hasha [17] and conducted a series of experiments in order to investigate characteristics of the produced sheet by the colliding jets containing a viscous fluid. They found that lots of geometrical and hydrodynamic parameters such as the impingement angle ($\alpha$), Weber number ($W_e$), and Reynolds number ($Re$) considerably affect the sheet thickness. In addition, they have presented an experimentally improved model for the two-jet impingement systems in which energy dissipation assumption is also considered. On the whole, they claimed that Hasson and Peck's expression [7] has the most accurate prediction for the liquid sheet thickness.

Many other experimental and numerical studies have been done recently to address other necessary assumptions [21, 22] which are missed by Ranz [13], Miller [15], and Hasson and Peck [7]. According to these studies, two principal and unanswered questions have been raised in associated with these expressions for the liquid sheet thickness; first, what is the theoretical reason that triggers the non-unique exact solutions, and second, which of these theoretical expressions has the most physical solution. The main object of the current study is answering these two questions by considering the minimum energy principle.

In classical physics, the principles of minimum energy and maximum entropy are two restatements for the second law of thermodynamics. The principle of minimum energy states that for a closed system with a constant entropy, the energy will decrease and approach to a minimum value at the equilibrium. By considering the formed sheet and bounded rim as a closed system, the most physical solution for the sheet thickness is the one that satisfies the minimum stored energy.

Here, it is proved that the non-unique solutions for the sheet thickness originate from the incompleteness of the governing equations. Then, as a new approach, the minimum energy principle is coupled with the conservative equations to attain the complementary set of equations. Based on the minimum energy principle the stored energy inside the sheet should be quantified to recognize the minimum one. The sheet area is calculated as the equivalent quantity instead of stored energy because of the inviscid assumption. To do this, the liquid sheet area is presented in terms of the sheet thickness and numerically solved using Runge-Kutta fourth-order method for the equations of Ranz [13], Miller [15], and Hasson and Peck [7]. The numerical calculations are repeated for three different impingement angles of 58, 89, and 117 degrees and compared. The comparison is shown that Hasson and Peck's expression is the most physical solution in all three impingement angles since it has the minimum area thus the least stored energy. Finally, an experimental study is conducted to visualize the liquid sheet shape using the Shadowgraph technique and validate the numerical results.

## Governing equations

A thin liquid sheet bounded by a rim is shown in Fig. 1a. that is initiated from the impingement point of two identical jets with the relative angle of $2\alpha$. The impingement point is shown by the red dot. In Fig. 1b, the schematic of the control volume is presented in which demonstrates the sheet is bounded by a rim.

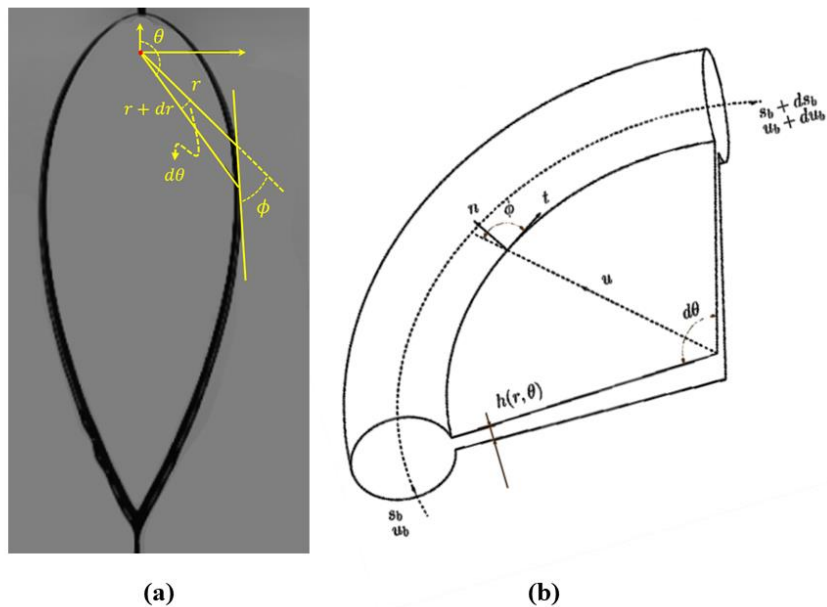

*Figure 1:* (a) The coordinate system for analyzing the liquid sheet formed by two identical impingement jets. (b) A Schematic for the assumed differential control volume that is used for deriving mass and momentum conservations.

Therefore, the mass conservation equation for the presented control volume could be written as:

$$2\rho \left(\frac{\pi d_0^2}{4}\right) u_0 = \int_0^{2\pi} \rho h u r d\theta \tag{1}$$

where, $d_0$ is the jet orifice diameter, $u_0$ is the jet velocity before the impingement point, $u$ is the fluid velocity through the liquid sheet, $h$ is the sheet thickness, $\rho$ is the fluid density, $r$ is the radial coordinate, and $\theta$ is the angular coordinate on the liquid sheet. Besides, the momentum conservation in the longitudinal direction of the liquid sheet could be expressed as below:

$$2\rho \left(\frac{\pi d_0^2}{4}\right) u_0^2 \cos\alpha = \int_0^{2\pi} \rho h u^2 \cos\theta d\theta \tag{2}$$

where, $\alpha$ is the half of the impingement angle. Since the flow is assumed inviscid, the flow velocity through the sheet ($u$) is constant as well as the velocity of the incoming jets. Three theoretical solutions for the sheet thickness by Ranz ($h_r$) [13], Miller ($h_m$) [15], and Hasson and Peck ($h_h$) [7] have been introduced in Eqs. 3-5. All these three expressions are valid simultaneously and satisfy both Eqs. 1 and 2.

$$h_r = \frac{d_0^2}{4r}(1 + 2\cos\alpha \cos\theta) \tag{3}$$

$$h_m = \frac{d_0^2 \sin^2\alpha}{4r(1 - 2\cos\alpha \cos\theta + \cos^2\alpha)} \tag{4}$$

$$h_h = \frac{d_0^2 \sin^3\alpha}{4r(1 - \cos\alpha \cos\theta)^2} \tag{5}$$

Obviously, only one of these three expressions could be the exact solution of Eqs. 1 and 2 and the others are non-physical solutions for the sheet thickness. The main focus of this work is presenting a meaningful theoretical explanation for rejecting non-physical solutions. In theory, the incompleteness of the governing equations and/or the assumptions are two common reasons for having non-unique solutions. As is already known, the complete set of governing equations includes mass, momentum, and energy equations. By assuming a potential flow for the sheet, the energy conservation equation in the liquid sheet could be added by applying Bernoulli's equation. Since the sheet is bounded by the rim, the behavior of the rim could be correctly used in our analysis. Therefore, using the control volume presented in Fig. 1a, the mass equation for the rim could be written as:

$$\rho \frac{d(u_b s_b)}{ds} = \rho u h \sin\varphi \tag{6}$$

where, $s_b$ is the rim cross-section area, $u_b$ is the rim-flow velocity, and $\varphi$ is the radial and tangential directions relative angle that is shown in Fig. 1a. Then, the rim momentum balance in the tangential and radial directions could be defined as Eqs. 7 and 8,

$$\rho \frac{d(u_b^2 s_b)}{ds} = \rho u^2 h \sin\varphi \cos\varphi \tag{7}$$

$$\rho u_b^2 s_b (d\theta + d\varphi) = 2\sigma ds - \rho u^2 h \sin^2\varphi \, ds \tag{8}$$

where, $\sigma$ is the surface tension coefficient. Moreover, the trigonometrical equations could be defined as below.

$$\sin\varphi = \frac{r d\theta}{ds} \tag{9}$$

$$\tan\varphi = \frac{r d\theta}{dr} \tag{10}$$

Based on the reference characteristics of the jets, $d_0$ and $u_0$, the nondimensionalized equations could be written as,

$$\frac{d(\tilde{u}_b \tilde{s}_b)}{d\theta} = F(\theta, \alpha) \tag{11}$$

$$\frac{d(\tilde{u}_b^2 \tilde{s}_b)}{d\theta} = F(\theta, \alpha)\cos\varphi \tag{12}$$

$$\tilde{u}_b^2 \tilde{s}_b \left(1 + \frac{d\varphi}{d\theta}\right) \sin\varphi = \frac{2\tilde{r}}{We} - F(\theta, \alpha) \sin^2 \varphi \tag{13}$$

$$\frac{d\tilde{r}}{\tilde{r}} = \frac{d\theta}{\tan\varphi} \tag{14}$$

where, $F(\theta, \alpha) = h(r, \theta, \alpha) \times r$ and $We$ is the Weber number, defined as $\rho u_0^2 d_0/\sigma$. In addition, $\tilde{r}$, $\tilde{s}_b$, and $\tilde{u}_b$ are the nondimensionalized forms of $r$, $s_b$, and $u_b$, respectively. By balancing the normal forces in Eq. 13, the normal momentum of the jet can be canceled out by the surface tension. From the energy conservation viewpoint, part of the kinetic energy is stored in the sheet and the rim that is also known as the surface or interfacial energy. This interfacial energy at the gas-liquid interface is reserved due to the surface tension. The surface tension acting as a resistance force that prevents more expansion of the liquid sheet. According to the minimum energy principle for a constant entropy state, the liquid holds a shape in which the minimum amount of energy, $E_s$, has been stored.

As is shown in Eq. 15, the stored energy could be calculated by multiplying the surface tension, $\sigma$, and the total area, $A$, that is sum of the liquid sheet and the rim surface areas. The integral form for calculating the total area is presented in Eq. 16. In this equation, the first integral accounts for the sheet surface area, the second integral is attributed to the rim lateral surface area, $r(\theta)$ is the radius of the liquid sheet at the rim edge, and $P_b$ is the local perimeter of the rim. The total area is calculated along the rim from 0 to L that is the rim length in the half side of the liquid sheet.

$$E_s = \sigma \times A \tag{15}$$

$$A = 2\int_0^\pi \int_0^{r(\theta)} \rho \, d\rho \, d\theta + 2\int_0^L \{P_b - h(r(\theta))\} \, ds \tag{16}$$

It is hypothesized that the total area (Eq. 16) is the crucial criterion by which the non-physical solutions would be determined. By linking the total area and the sheet thickness, the most physical solution is the one by which the minimum total area resulted and the minimum energy principle satisfied. Clearly, the non-physical solutions predict a wider shape for the liquid sheet. Mathematically, the hypothesized model can be expressed as Eq. 17.

$$h_{physical}(r,\theta,\varphi) = \min_i A(h_i) \qquad (17)$$

There is an implicit coupling between the shape of the liquid sheet and its thickness but it doesn't an analytical solution. So, a numerical solution should be applied to predict the liquid shape. The thickness expression with the minimum surface area satisfies the second law of thermodynamics and theoretically could be introduced as the physical solution for the sheet thickness. This should be noted that the isothermal and inviscid flow assumptions are not considered. Therefore, the second law of thermodynamics might need to be applied implicitly in the equations.

## Results and discussion

In this section, our hypothesis is tested by calculating the liquid sheet shape and surface area. At first, Eqs. 11-14 are solved numerically to obtain the shape of the liquid sheet for each of the thicknesses presented at Eqs. 3-5. Then, the stored interfacial energy is calculated based on Eqs. 15 and 16. Finally, the results are compared to find the most compatible liquid sheet thickness with the minimum stored energy, also called the physical solution. Indeed, all of the three expressions cannot be true for a unique physical phenomenon and the minimum energy principle is considered to determine non-physical solutions.

The first-order nonlinear differential equations in Eqs. 11-14 are numerically solved by Runge-Kutta fourth-order method. The initial conditions are defined at $\theta = 0$, where the initial radius of the rim is slightly back from the stagnation-point, $\tilde{r}(0) = \tilde{r}_0(1-\epsilon)$, $\epsilon = 0.01$, and $\tilde{r}_0 = We \sin^3 \alpha [8(1+\cos^3 \alpha)^2]$. In addition, $\varphi(0) = \sin^{-1}\sqrt{1-\epsilon}$ and $\tilde{u}_b(0) = \cos\varphi = \epsilon$ are applied for Eqs. 11-13. In Eq. 11, the initial condition for $\tilde{s}_b$ is estimated using the experimental data [23]. The numerical solutions for the sheet shape in three different impinging angles of $58^o$, $89^o$, and $117^o$ is demonstrated in Fig. 2. In this figure, the plots of a, b, and c are respectively obtained by Ranz [13], Miller [15], and Hasson and Peck [7] expressions. Due to the sharp edge, it is visually obvious that Ranz's theory [13] predicts the most non-physical shape for the liquid sheet. For the first time, this prediction deviation with the experimental results has been reported by Miller [15].

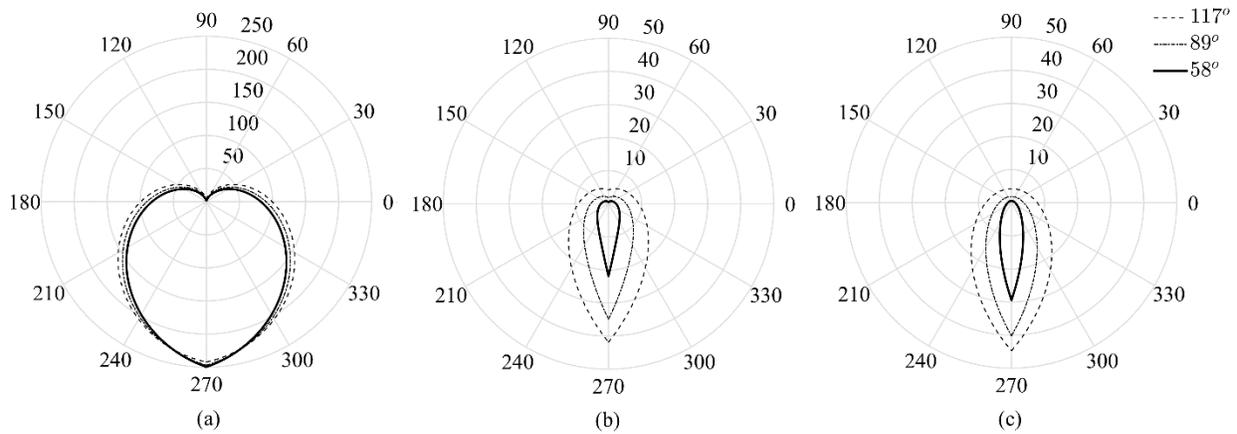

*Figure 2:* *The numerical results for the sheet shape in three different impingement angles of $58^o$, $89^o$, and $117^o$ using the sheet thickness expressions that are presented by (a) Ranz [13], (b) Miller [15], and (c) Hasson and Peck [7].*

However, Ranz's theory could be rejected visually, the difference between the results of the liquid sheet shapes in Fig.2b and c are indistinguishable. Therefore, an experimental study is needed to closely check the difference. The schematic of the experimental setup is illustrated in Fig. 3. In this setup, a centrifugal pump conducts the liquid from a tank to a high-pressure vessel, and then a nitrogen gas vessel pressurizes the high-pressure container to particular uniform pressure. The flowmeters with the precision of 0.2 mL/min measure the jets' volume flow rates that are produced with two stainless steel hypodermic needles. These needles have square ends and an internal diameter ($D_j$) of 0.6 mm. The impinging angle is fixed at 90 degrees, utilizing a rotary stage with the precision of 1 degree. A micro-stage with an accuracy of 10 $\mu m$ is also used to ensure the two impinging jets are perfectly aligned. A mixture of Glycerol-water with the viscosity of $\mu = 0.0155\ Pa.s$ and surface tension of $\sigma = 0.062\ N/m^3$ is used as the working liquid. The results are obtained by the Shadowgraph technique and the liquid sheet shapes are visualized in different impingement angles. A pulsing light source called PhotoFreezer v2.5" with pulse time down to 125 nanoseconds, and frequency up to $1000\ Hz$ is synced with a high-speed CCD camera (ArsinTNF) that could capture images up to 300 frames per second.

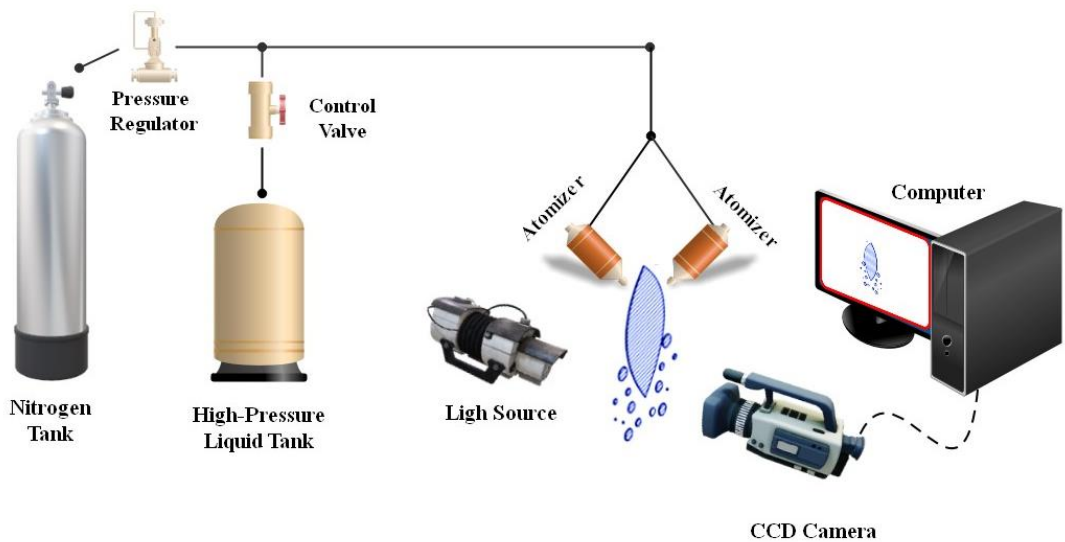

*Figure 3: The experimental setup used for the liquid sheet visualization based on the Shadowgraph technique.*

In Fig. 4, the experimental results are compared with the numerical liquid sheet shapes predictions using Miller [15], and Hasson and Peck [7] expressions. The experiment is done for a specific condition in which two jets are impinging with constant and identical velocities of $3.3\ m/s$ and the relative angle of 90 degree. This figure reveals that Hasson and Peck's theory predicts a more real shape compared with the Miller's theory.

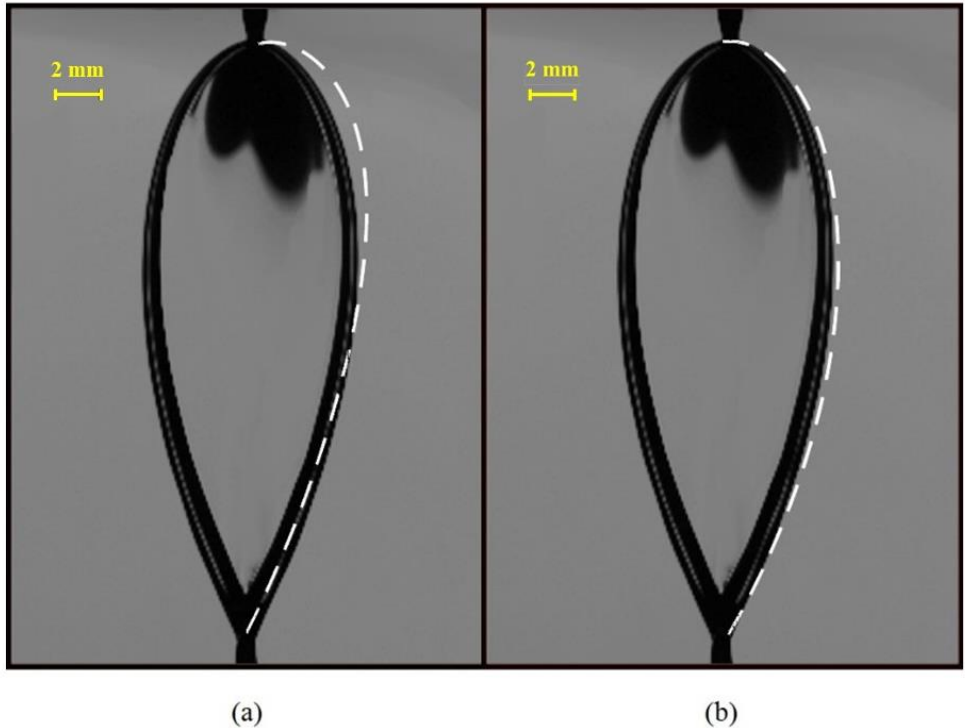

*Figure 4:* *Experimental and numerical results comparison. The sheet shape is visualized with the Shadowgraph technique and the liquid sheet shapes are obtained numerically using (a) Miller's [15] and (b) Hasson and Peck's [7] expressions.*

In the final analysis, the results of the lateral surface areas are compared for these 3 expressions and presented in Table 1. The lateral surface of the liquid sheet for three different impinging angles of $58^o$, $89^o$, and $117^o$ is numerically calculated according to the thickness presented in Eqs. 3-5. According to the energy equation, the minimum level of stored interfacial energy is attributed to the minimum lateral surface of the liquid sheet as well as, the more physical thickness [20]. Comparing the results of Table 1 proves that the minimum lateral surface is attributed to Hasson and Peck's theory [7] and this theory is the most physical one. As is expected from our numerical results and Miller's experiments [15], Ranz's theory predicts an extremely high lateral surface area and it could be attributed to the most irrelevant physical conditions. In addition, these calculations show that Hasson and Peck's expression is the most physical solution for the sheet thickness. Therefore, the minimum energy principle is a crucial constraint that unifies the solution of Eqs. 1 and 2.

*Table 1: Non-dimensional lateral surface area of the liquid sheet calculated for three impinging angles using Ranz, Miller, Hasson and Peck sheet thickness equations.*

|  | Non-dimensional surface area ($A$) | | |
| --- | --- | --- | --- |
| Impingement angle | $58^o$ | $89^o$ | $117^o$ |
| Ranz [13] | 305993 | 289142 | 273347 |
| Miller [15] | 5110 | 2847 | 1025 |
| Hasson and Peck [7] | 4795 | 2386 | 673 |

## Conclusion

To date, a convincing theoretical reason has not been presented to reject the incompatible solutions. In the present work, the minimum energy principle was applied to reject the liquid sheet thickness non-physical solutions that satisfy conservation equations. In addition, a criterion was introduced to find the most physical solution. First, it was assumed that the liquid sheet bounded with a liquid rim, and the governing equations were derived. In addition, it was hypothesized that the minimum energy principle should be considered for rejecting non-physical solutions. To quantify the liquid sheet stored energy, the liquid lateral surface area was calculated by the numerical prediction of the sheet shape. Since the inviscid flow assumption was reasonably assumed, the minimum energy principle satisfies when the liquid sheet has the minimum lateral surface area.

To calculate the lateral surface area, the sheet shape was derived as a function of the sheet thickness, then numerically solved for three well-known expressions for the sheet thickness presented by Ranz [12], Miller [14], and Hasson and Peck [6]. The corresponding lateral areas were numerically calculated then compared at three impingement angles of 58, 89, and 117 degrees, exclusively. The numerical results in all three impinging angles proved that Hasson and Peck's expression is the most physical solution for the sheet thickness since its predicted liquid shape had the minimum lateral surface area, therefore the minimum amount of the stored energy. The numerical comparison also proved that Ranz's expression had the maximum stored energy in all three conditions. Therefore, it was presented as the most non-physical solution for the sheet thickness.

To be more confident about the numerical comparison, an experimental study was performed using the Shadowgraph technique to visualize the sheet shape. The numerical sheet shapes were compared with the experimental results and it was concluded again that Hasson and Peck's expression is the most reliable solution for the sheet thickness. In all, it was numerically and experimentally demonstrated that the minimum energy principle is the overlooked supplementary constraint by which the unique physical solution would be obtained.

## Acknowledgment

The authors thank Arsin Tabesh Negran Fannavar Company for their sincere cooperation and the availability of PhotoFreezer v2.5. The financial support from the center of research assistance of Sharif University of Technology (Grant number G950411) is also gratefully acknowledged.